# Dynamic wetting and heat transfer during droplet impact on bi-phobic wettability-patterned surfaces

Wenliang Qi(齐文亮),[1] and Patricia B. Weisensee,[1, 2, a)]

**AFFILIATIONS**

[1] Department of Mechanical Engineering and Materials Science, Washington University in St. Louis, St. Louis, Missouri 63130, USA

[2] Institute of Materials Science and Engineering, Washington University in St. Louis, St. Louis, Missouri 63130, USA

[a)] Author to whom correspondence should be addressed: p.weisensee@wustl.edu

**ABSTACT**

This paper reports the dynamic wetting behavior and heat transfer characteristics for impinging droplets on heated bi-phobic surfaces (superhydrophobic matrix with hydrophobic spots). A non-patterned superhydrophobic and a sticky hydrophobic surface acted as control wettability surfaces. As expected, differences in wetting and heat transfer dynamics were noticeable for all surfaces, with the most pronounced variation during the receding phase. During spreading, inertia from the impact dominated the droplet dynamics and heat transfer was dominated by convection at the contact line and internal flow. As contact line velocities decreased over time, evaporative cooling at the contact line gained importance, especially for the bi-phobic surfaces, where liquid remained trapped on the hydrophobic spots during receding. These satellite droplets increased the contact area and contact line length, and assisted heat transfer and substrate cooling after lift-off of the main droplet. Compared with the hydrophobic surface, the contribution of the contact line heat transfer increased by 17 to 27% on the bi-phobic surfaces, depending on the location of impact relative to the hydrophobic spots. Nonetheless, the bi-phobic surfaces had a lower total thermal energy transfer. However, compared with the plain superhydrophobic surface, heat transfer was enhanced by 33% to 46% by patterning the surface. Depending on the application, a trade-off exists between the different surfaces: the sticky hydrophobic surface provides the best cooling efficiency, yet is prone to flooding, whereas the superhydrophobic surface repels the liquid, but has poor cooling efficiency. The bi-phobic surfaces provide a middle path with reasonable cooling effectiveness and low flooding probability.

## I. INTRODUCTION

Water droplet impact dynamics on isothermal superhydrophobic surfaces have been extensively studied over the last decade.[1] With the advent of high-speed thermal imaging, there has been significant interest in determining local heat transfer rates during water droplet impact on non-isothermal superhydrophobic surfaces.[2–5] When droplets impact hot or cold surfaces, the resulting heat transfer and potential phase change can alter the fluid dynamics compared to adiabatic impact.[6] Many experimental studies have focused on characterizing impact dynamics on hot surfaces, and developed regime maps for convection, film evaporation, nucleate boiling, and film boiling (Leidenfrost) as a function of substrate temperature and droplet impact velocity.[7–14] Detailed understanding of temperature profiles and evaporation rates is critical for spray cooling[15] or internal combustion engines.[3] For millimetric droplets impacting a superhydrophobic surface below the Leidenfrost temperature, only a small fraction of the total available thermal energy is transferred due to a very short contact time (~ ms). Smaller droplets transfer a larger fraction of their total thermal energy despite having a shorter contact time.[2,4] This increase in cooling effectiveness was attributed to the decreased thermal capacitance and a larger effective heat flux due to enhanced conduction through small droplets. On hydrophilic and hydrophobic surfaces, increasing the impact Weber number (*i.e.*, impact velocity) also increases the cooling effectiveness due to an increased droplet-substrate contact area and higher internal droplet velocities.[15,16] However, on a superhydrophobic surface, the impact velocity has only a minor influence on heat transfer and fluid flow due to the contact time invariance.[17,18] For all substrate wettabilities, the highest heat transfer rates are typically associated with forced convection, as well as evaporation in the contact line region during droplet spreading.[6,19–21] Vortices within the droplet transport hotter fluid from the central region to the liquid-gas interface, promoting higher evaporation and sensible heat transfer near the three-phase contact line (TPCL). During receding, local heat transfer maxima are lower than during spreading, but remain located near the contact line.

In recent years, researchers have studied wettability-patterned surfaces and have demonstrated their great potential for industrial applications, such as directional and passive liquid transport, targeted deposition, and enhanced phase change heat transfer and water harvesting.[22–36] Wettability-patterned surfaces usually contain co-located areas of two or more different wettabilities (superhydrophobic, hydrophobic, hydrophilic, or superhydrophilic), helping to achieve the desired wetting properties and to allow the control of heat transfer.[36–39] For example, Koukoravas et al.[39] showed that diverging superhydrophilic tracks on a superhydrophobic background can transport impinging jets and enable spatial control of cooling while simultaneously enhancing evaporation heat transfer. From a hydrodynamic perspective, when droplets impact a chemically patterned surface, the spreading dynamics are similar to those on a surface with single wettability. However, droplets can split during receding and be divided into satellite droplets due to modified hydrodynamics and contact line pinning at the wettability boundary.[40–42] Diby et al.[43] reported that an initially static water droplet sitting on the interface between superhydrophobic and hydrophilic regions on a surface moves towards the hydrophilic region. Yang et al.[44] found that during droplet impact, the direction of rebound and distribution of splashing satellite droplets can be regulated by the shape and size of the wettability pattern on a hybrid superhydrophobic/-philic surface. Given that surface wettability strongly affects heat transfer during droplet impingement,[17,45] it is to be expected that wettability-patterned surfaces can be used to manipulate and control the thermal signature of heated surfaces during droplet impact.

To overcome the time-constraint for heat transfer on superhydrophobic surfaces, while simultaneously mitigating substrate flooding from droplet deposition, we propose the use of bi-phobic wettability-patterned (super)hydrophobic surfaces. The focus of the present research is to investigate droplet wetting and heat transfer behavior on heated superhydrophobic (SHP), hydrophobic (HP), and bi-phobic (superhydrophobic matrix with hydrophobic patterns) surfaces for different surface temperatures and impact heights. Using synchronized high-speed optical and infrared (IR) imaging, we are able to correlate droplet dynamics to the spatial distribution of the solid-liquid interfacial temperature, heat flux, and the total heat transfer to the droplet. Furthermore, we elucidate the influence of surface wettability and contact line dynamics on convection and evaporation heat transfer. Findings from this work have the potential to improve the efficiency of spray cooling applications,[12] or of micro-reactors in chemical and biomedical engineering.[46,47]

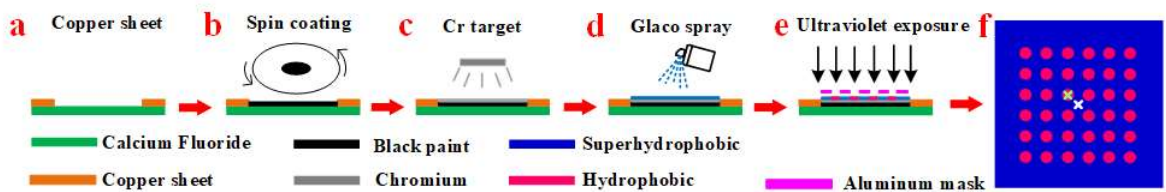

FIG. 1. Schematic of the heater design and fabrication process of the bi-phobic surface. The white and green crosses in (f) show the locations for "impact center" (IC; white x) and "impact dot" (ID; green x), respectively.

## II. EXPERIMENTAL METHODS

### A. Heater design and surface processing

The fabrication process, the heater design, and the corresponding wettability patterns are shown in Fig. 1. First, two copper sheets were attached to opposing sides of an IR-transparent calcium fluoride substrate (UQG Optics, thickness 1.18 mm). Then a 12 μm thin layer of black paint (Testors 18PK, Black Eenamel, 1149TT), which serves as transducer for the thermal imaging, was spin-coated for 90 s at 1200 rpm onto the CaF2 substrate. The thickness of the black paint layer was measured using a profilometer (Alpha-Step D-100 Stylus Profiler). On top of the black paint layer, a 350nm thin chromium film was deposited using physical vapor deposition, serving as heater (effective heater area: 16.5 mm × 25.5 mm). The two copper sheets served as voltage transducers between the power supply (Instek PSW 160-7.2) and the Cr heater to achieve uniform heating. Lastly, the commercially available superhydrophobic Glaco

Mirror coat was sprayed on the chromium layer and dried for a minimum of 24 hours.[48] A superhydrophobic surface with static contact angle $\theta$ = 162° ± 2° was obtained. To selectively wettability-pattern the surfaces, the heater assembly was exposed to ultraviolet radiation (PSD Pro Series Digital UV Ozone System) with a patterned aluminum mask (circular holes of 0.8 mm in diameter and a pitch of 1.6 mm). After 105 minutes of exposure, the bi-phobic surface was obtained: exposed regions turned hydrophobic (static contact angle of ~ 100°), while the unexposed regions remained superhydrophobic. Using scanning electron microscopy (SEM) images, we confirm that the surface topography remained largely unchanged during UV exposure. Only the nanoscale roughness increased slightly due to the removal of some polymer, as can be seen in the SEM micrographs in Fig. 2. Despite a relatively high static contact angle of approximately 100° on the exposed surface, water fully wetted the nanoroughness in a Wenzel state with a receding contact angle approaching zero. We hence call this exposed area "sticky hydrophobic" (often also referred to a Rose petal effect). To distinguish the impact location with respect to the surface patterning, we refer to droplets impacting centrally in-between hydrophobic dots on the superhydrophobic matrix as "impact center" (IC) and impacts centered on a hydrophobic spot as "impact dot" (ID), as represented by the white and green markers, respectively, in Fig. 1.

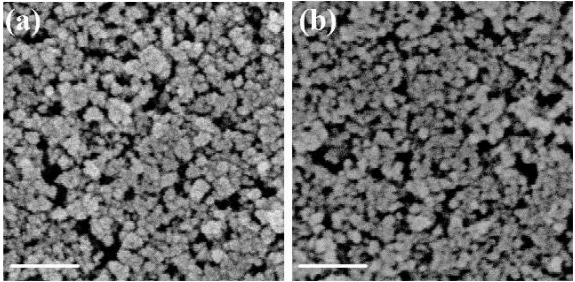

FIG. 2. Scanning electron microscopy (SEM) images of the Glaco layer on a glass substrate (a) before and (b) after UV treatment. Scale bars are 250 nm.

**B. Experimental setup**

A schematic of the experimental setup is shown in Fig. 3. Side-view images of the droplets were recorded at 10,000 frames per second (fps) using a Photron Mini AX200 high-speed camera with a Canon MP-E 65mm f/2.8 1-5X Macro Lens at a spatial resolution of 3.3 μm/pixel. A Telops FAST M3k high-speed mid-wave infrared (IR) camera, equipped with a 1x long working-distance lens (Telops), captured the temperature distribution of the samples in bottom-view (using a Thorlabs gold-coated mirror) at 5000 fps with a spatial resolution of 30 μm/pixel. The IR transparency of the $CaF_2$ allowed to measure the temperature distribution on the upper surface (*i.e.*, the droplet-solid interfacial temperature). Droplets of de-ionized (DI) water were generated at the tip of a gauge 30 needle, connected to a syringe pump (New Era NE-1000). Droplets detached due to gravity with diameters $D$ = 2.23±0.05 mm. Droplets impacted the bi-phobic surfaces from heights of 80±1 mm, leading to an impact velocity of 0.74 m/s (Weber number $We \approx 10$). Using resistive (or Joule) heating in the Cr thin film, the substrates were heated to 75 ± 0.5°C prior to impact. These substrate temperatures were well below the Leidenfrost point to prevent nucleate or film boiling of the impinging droplets. The substrate temperature was chosen to optimize for an acceptable signal-to-noise ratio during thermal imaging, while avoiding the deterioration of the bi-phobic polymer coating. Through preliminary testing, the necessary voltage across the thin film was determined to achieve the desired substrate temperature, where Joule heating balanced convective losses to the air. Prior to each experiment, the sample was given sufficient time to reach steady state. The uniformity of the heat flux on the heater surface was confirmed by observing a homogeneous temperature distribution prior to droplet impact, as seen in Fig. 4b at $t$ = 0 ms. During the experiments, the voltage was then kept constant. Given that the change in electrical resistance with temperature is minimal within the herein observed temperature ranges (few K), the volumetric heat generation in the Cr thin film is expected to be constant throughout the duration of each experiment.

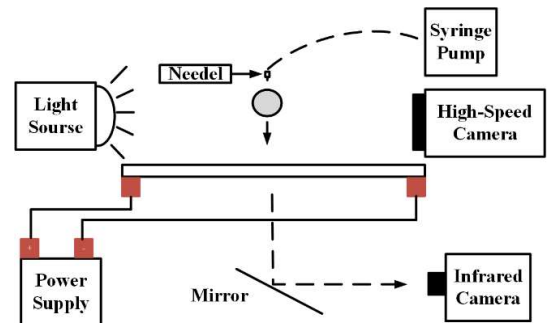

FIG. 3. Schematic of the experimental setup.

**C. Heat transfer analysis**

In order to obtain heat transfer rates from the temperature distribution at the substrate-droplet interface, we followed the same procedure as in our previous work.[49] Briefly, a transient energy balance was applied to each pixel element,[48,50] where the convective heat flux of each pixel element is expressed as:

$$q''_{Top} = q_J/A - Q_{Store}/(\Delta t \cdot B) - q''_{Side} - q''_{Bottom}, \quad (1)$$

where $q''_{Top}$ is the heat flux to the top surface (*i.e.*, droplet or air) of each pixel element, $q_J$ is the heat generated in the Cr thin film heater by Joule heating, $A$ is the effective area of the heater, $Q_{Store}$ is the thermal energy stored in the pixel element in the time interval $\Delta t$ between two successive frames, $B$ is the area of a pixel element, $q''_{Side}$ is the heat flow due to conduction to/from the four neighboring pixels, and $q''_{Bottom}$ is the heat loss to the environment through the bottom of the calcium fluoride window. The individual terms are expressed as:

$$q_J = I_s^2 R_{Cr}, \quad (2)$$

$$Q_{Store} = B^2(\rho_{Cr}C_{Cr}\delta_{Cr} + \rho_{BP}C_{BP}\delta_{BP})(T_{x,y,\tau+1} - T_{x,y,\tau}), \quad (3)$$

$$q''_{Side} = -(k_{Cr}\delta_{Cr} + k_{Bp}\delta_{Bp})\left(\frac{\partial^2 T_s}{\partial x^2} + \frac{\partial^2 T_s}{\partial y^2}\right), \quad (4)$$

$$q''_{Bottom} = \left(\frac{1}{h_{air}} + \frac{\delta_{CF}}{k_{CF}}\right)^{-1}(T_s - T_{air}). \quad (5)$$

where $I$ is the electrical current, $R$ is the electrical resistance, $\rho$ is the density, $C$ is the specific heat capacity, $\delta$ is the layer thickness, $k$ is the thermal conductivity, $T$ is the temperature, and $h$ is the convective heat transfer coefficient. The subscripts are defined as follows: $\tau$ is the time step between successive frames, $x$ and $y$ are Cartesian coordinates, $Cr$ represent the chromium layer, $BP$ the black paint, $s$ is the heater surface, $CF$ stands for the calcium fluoride substrate, and $air$ signifies the ambient air.

### D. Data analysis and processing

The recorded thermal images were processed using a custom written MATLAB code to obtain temporal heat flux distributions. In order to account for non-ideal substrate properties in the infrared spectrum (*e.g.*, black paint emissivity, calcium fluoride transmissivity, mirror reflectivity, and the thermal resistance of the Glaco coating), all temperatures readings were calibrated using a correction correlation that relates the IR camera readout to an RTD surface measurement (Omega, PT100) at the top side of the sample. To obtain the electrical resistance of the Cr layer for each individual sample, an energy balance was applied prior to droplet impact. The energy balance contains natural convection and thermal radiation exchange at the upper surface of the sample ($q''_{Top}$) (assumption: emissivity Glaco $\varepsilon_r$ = 1) and all other terms discussed in Eq. (1). Along with the measured electrical current, we determined a total heater resistance of 8-10 Ω for all samples, in line with resistivity data for sputtered or e-beam deposited thin film Cr.[51] Applying the principle of error propagation, the overall uncertainty in reported heat flux values is 4.4 to 11% (see section A in the SI).

All experiments were repeated at least three times. Figure S1 exemplarily compares average temperatures and total integrated heat fluxes for IC bi-phobic surfaces for the three impact events. The relative error between each trial is less than 10%, and hence falls within the overall experimental uncertainty. Additional experiments were conducted at impact heights of 40 and 120 mm and surface temperatures of 45°C and 60°C, for which results are shown in the Supporting Information (SI). The general conclusions drawn in this manuscript are consistent with those from the supplementary experiments.

## III. RESULTS AND DISCUSSION

### A. Overview of droplet dynamics and heat transfer

Figure 4 gives a high-level comparison of side-view images and corresponding bottom-view temperature and heat flux distributions for superhydrophobic (SHP), hydrophobic (HP), and bi-phobic surfaces (ID and IC). Individual observations will be discussed in more detail in the following sections of this manuscript. It can be clearly seen that the initial droplet spreading, which is dominated by inertia, is similar on all surfaces (also compare to Fig. S2).[52] Unlike for droplet impact on a heated surface with a temperature above the boiling point of water,[53,54] we observe a distinct annular pattern in the temperature and heat flux profiles during spreading. We believe this is the first report to directly visualize the influence of the well-studied droplet hydrodynamics on the thermal signature with such spatial resolution.[5,55–59] We can clearly distinguish the rim from the rest of the spreading lamella.[60] On the superhydrophobic and bi-phobic surfaces, we observe two local maxima in the heat fluxes: one narrow region near the spreading three-phase contact line, and one narrow region inside the droplet. During spreading, vortex rings are generated within the droplet and the rim. This internal flow results in locally varying heat transfer coefficients, which lead to the distinct circular heat flux pattern, along with irregular liquid supply from the bulk to the interface.

A similar phenomenon was also observed by Herbert *et al.*,[19] who concluded that the secondary local maximum indicates a region where the film thickness exhibits a minimum, and hence a low thermal resistance. From Fig. 4 we also see that in contrast to the non-wetting surfaces, the heat flux distribution on the sticky hydrophobic surface is more uniform and of higher magnitude, and heat flux maxima do not necessarily occur near the contact line region, but within the center region of the droplet, especially for higher impact velocities.[5] Some of these features become more obvious at higher impact velocities, as shown in Fig. S3 for an impact height of 120 mm (0.74 m/s). We attribute this behavior to two fluidic phenomena. First, at times after maximum spreading (> 2.8 ms), we can clearly identify regions where colder water from the bulk of the droplet is transported towards the interface due to droplet oscillations and the aforementioned internal flow.[61] Second, due to the Wenzel state of the droplet on the rough hydrophobic surface, the actual droplet-substrate interfacial area is increased substantially compared to the superhydrophobic case or even a smooth hydrophobic surface. This increased contact area leads to a more efficient heat transfer between the substrate and the droplet. Based on the same concept, we observe higher local heat transfer rates near the hydrophobic regions on the bi-phobic surfaces, once the droplet passes over these during spreading. At later times, as spreading velocities decrease, as quantified in Fig. S4, this region of enhanced contact becomes more obvious. As the influence of droplet impact inertia and convection decrease, evaporation at the three-phase contact line becomes progressively important. At maximum spreading and during receding, the highest heat flux occurs predominantly near the TPCL for the non-wetting surfaces. On the superhydrophobic surface, the impacting droplet spreads to its maximum diameter, recedes, and bounces off from the surface at its capillary-inertial time.[18] After lift-off, the temperature at the center of the impact location remains lower than the rest of the sample for a short period of time, likely due to thermal lag, whereas the heat transfer vanishes.[17,53] On the bi-phobic surfaces, due to energy barriers and contact line pinning at the wettability boundaries, the receding contact line distorts and loses its symmetric shape. Small droplets break off from the bulk of the receding droplet and deposit as satellite droplets on the hydrophobic spots. The very strong local heat transfer at the periphery of the hydrophobic spots indicates an evaporation-dominated heat transfer. On the bi-phobic surfaces, the phase change of these small satellite droplets contributes significantly to the overall cooling efficiency of the system. It is evident that the location of impact is very important for the dynamics of wetting and heat transfer during droplet impingement, as seen by comparison between ID and IC impact, and will be discussed in more detail later. On the rough hydrophobic surface, viscous dissipation is large and the droplet remains pinned to the surface instead of receding. Due to a considerably longer contact time and larger area of contact, the total heat transfer is increased and the substrate cooling effect is stronger and more prolonged.[17]

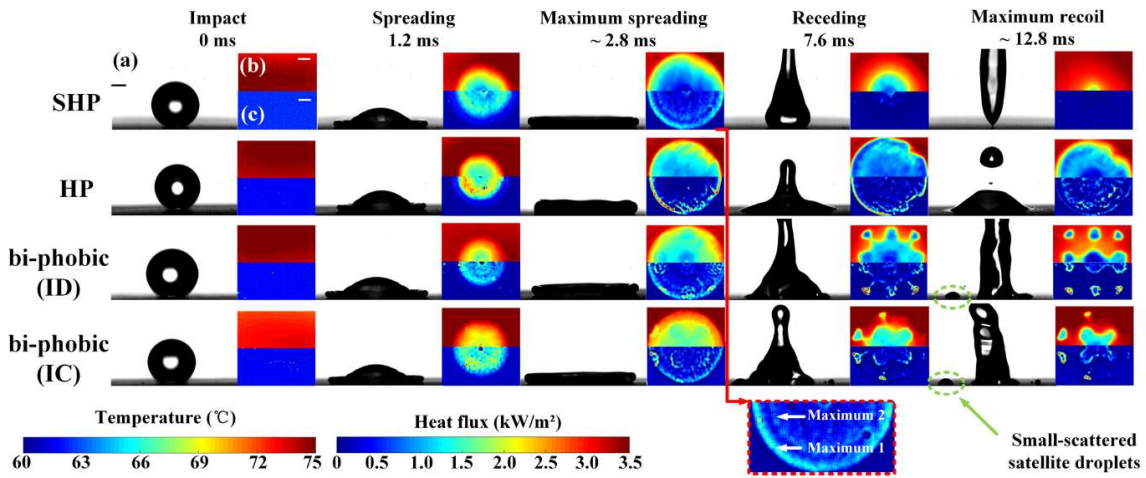

**FIG. 4.** (a) High-speed camera images of droplets and (b) corresponding temperature and (c) heat flux distributions for the different surfaces. Scale bars are 0.8 mm. SHP: superhydrophobic, HP: hydrophobic, ID: impact dot, IC: impact center. Videos in the supplementary material can be used for comparison of the temporal evolution of droplet dynamics and thermal signatures for different impact conditions.

### B. Comparison of thermal signatures

After establishing a framework on the fluid dynamics and heat transfer, we now turn to quantifying the thermal signatures of the different surfaces, as shown in Fig. 5. Specifically, we compare the average temperature $\bar{T}_s = \sum_{i=1}^{i=n} T_{s,i}/n$, the total heat flux $q''_{total} = \sum_{i=1}^{i=n} q''_{Top,i}/n$, and the total heat transfer $Q(t) = \int_0^t q''_{total} A(t^*) dt^*$, where $n$ is the number of pixel that encompasses a square with a side length equal to the maximum spreading diameter $D_{max}$. $A(t)$ is the time-dependent projected liquid-solid contact area, and is plotted in Fig. 6. We observe some general trends: First, the heat transfer performance is substantially different during the different phases of the drop impingement process (spreading and receding). Although the surface heat flux reaches its maximum value during spreading (Fig. 5b), the lowest temperatures (Fig. 5a) occur after reaching the maximum spreading diameter during the initial stages of receding (around $t \approx 4$ ms) for all surface types. This minimum in temperature is followed by a relatively long period of modest increase for superhydrophobic, ID and IC bi-phobic surfaces due to the decreasing contact area, which restricts the heat transfer between surface and droplet. The only exception to this trend is the hydrophobic surface, where the droplet deposits nears its maximum diameter and temperatures remain nearly constant. Similar to the temperature curves, the influence of surface patterning on heat flux rates (Fig. 5b) during spreading is negligible. The total heat flux reaches its highest value before maximum droplet spreading, and decrease quickly thereafter.[5,16] Most of the receding happens at a nearly constant heat flux, consistent with numerical investigations.[19] In this phase, the contact line moves over the region previously wetted and cooled, reducing the temperature difference necessary to achieve high heat transfer rates. Figure S5 shows that the same temperature and heat flux trends are also observed at different experimental conditions. The similarity in heat transfer rates leads to approximately the same amount of total thermal energy transfer during spreading. However, the relative amount of transferred heat during spreading and receding varies for the different wettabilities. For bi-phobic surfaces, the percentage of heat transferred during the spreading phase is approximately 50% of the total energy transfer, whereas spreading accounts for nearly 80% of the total heat transfer on the superhydrophobic surface (Fig. 5c). As expected, on the hydrophobic surface the total amount of heat absorbed by the droplet is greatest during the "receding" stage, due to a longer time period of contact and heat transfer (~10 ms vs. <3 ms).

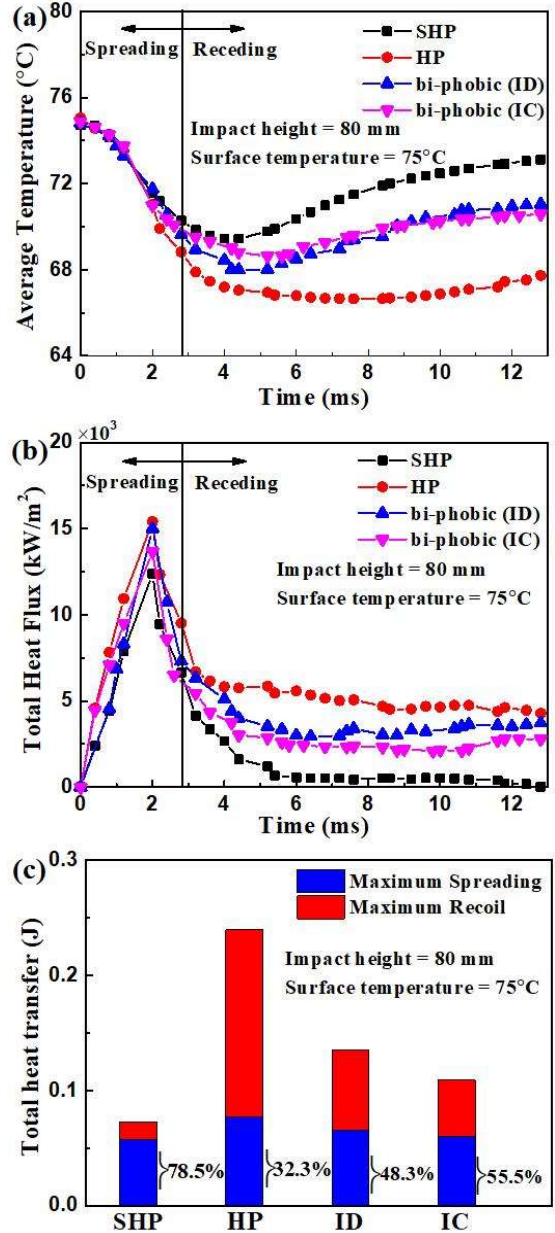

FIG. 5. Transient behavior of (a) average temperature, (b) total integrated heat flux and (c) total heat transfer for different surfaces within a total time period of 13 ms. The percentages shown in (c) represent the contribution of heat transfer during spreading over the 13 ms time frame.

This leads us to our second general observation, that surface wettabilities strongly influence the minimum temperature, the heat flux, and the total heat transfer. The lowest temperature and highest heat transfer rates are achieved with the sticky hydrophobic surface due to an enlarged contact area and pinning between droplet and substrate. During spreading, the surface roughness increases the effective contact area, and droplet deposition after spreading increases the time available for heat transfer to occur. On the other hand, droplets on the superhydrophobic surface with nearly the same roughness are cushioned by air pockets, reducing the effective contact area and heat transfer ability.[4,17] On the bi-phobic surfaces, the impact location dictates the fluid deposition during receding, and with that the overall cooling effectiveness of the impact event. Compared to the superhydrophobic surface, the satellite droplets increase the overall contact area and contact time, and consequently increase the total energy transfer by 33% (IC) and 46% (ID), respectively. The contribution of the receding phase to the total energy transfer dominates this increase, while the increase in energy transfer during spreading is only approximately 7%. When comparing these observations to Fig. 4, we see that the ID surface has a higher number of satellite droplets than IC, as some of the hydrophobic spots on IC were not able to pin the liquid during receding (for a different impact velocity, such as shown in Fig. S3, the number of satellite droplets can vary). These findings support our hypothesis that the formation of small-scattered satellite droplets strongly influences heat transfer dynamics through a higher surface area coverage and evaporation enhancement along the increased TPCL compared to droplets impinging a superhydrophobic surface.

### C. Influence of contact area

To quantify the influence of an increased surface area, in Fig. 6a we plot the time evolution of the apparent, *i.e.* projected, contact area, neglecting nanoscopic contact in the Wenzel state, as derived from the bottom-view IR images. From Figs. S2 and S4 we know that droplets have the same contact line velocity and spreading time on all surfaces, meaning that the contact area should be equal for all surfaces during spreading. However, as discussed above, during receding, the wettability of the surface strongly influences the apparent contact area of the droplets. The superhydrophobic surface shows the typical receding and lift-off behavior for a non-wetting surface.[62,63] In contrast, the droplet on the sticky hydrophobic surface behaves similarly to a wetting surface, where the droplet pins after reaching its maximum diameter. On the bi-phobic surfaces, during liquid retraction, some liquid pins on the hydrophobic regions, and the droplet breaks up into smaller satellite droplets that increase the contact area for bi-phobic surfaces compared to a superhydrophobic surface. As seen in Fig. 4, the ID impact configuration covers more of these hydrophobic spots, consistent with the higher wetting area observed in Fig. 6a. It is known that a larger contact area is helpful for heat transfer between surface and droplet.[20] Based on the transient behavior of the contact areas, we use the cooling effectiveness $\varepsilon(t)$ to analyze how well an impinging droplet cools a heated surface:[15]

$$\varepsilon(t) = \frac{Q(t)}{Q_{max}} = \frac{\int_0^t q''_{total} A(t^*) dt^*}{C_{p,d} M_d (T_{0,s} - T_{0,d})}. \quad (6)$$

Here, $C_{p,d}$ is the droplet specific heat capacity, $M_d$ is the droplet mass, and $T_{0,s}$ and $T_{0,d}$ are the temperatures of surface and droplet prior to impact, respectively. Guo *et al.*[2] found that more than 90% of the thermal transport to the droplet occurs during the spreading process on a superhydrophobic surface. After that, the cooling effectiveness tended to be constant. We observe a similar trend, as shown in Fig. 6b, and in line with Fig. 5c, where approximately 80% of the total thermal energy was transferred during spreading. However, the cooling effectiveness continuously increases in time for the other surfaces. During the spreading phase, all surfaces had a similar trend of cooling effectiveness, given that their spreading dynamics and heat transfer were similar during this phase. At later times, the cooling effectiveness on the hydrophobic surface continuously increases. Since $\varepsilon \sim A$, and $A \approx const.$ on the hydrophobic surface, whereas the contact area decreases on the other surfaces, the cooling effectiveness is distinctly higher for the hydrophobic surface for a single droplet. One question that naturally arises is whether the initial droplet size would influence this conclusion. For droplets larger than the ones measured here, the maximum cooling capacity would increase cubically ($Q_{max} \sim M_d \sim D_0^3$), whereas the area only scales quadratically with the droplet diameter ($A \sim D_0^2$). Hence, for larger droplets the cooling efficiency is expected to decrease.[2] On the other side of the spectrum, for microscopic droplets, past research has found that the spreading dynamics are similar to those of millimetric droplets.[64] Following the same argument as for larger droplets, this would entail an increased cooling efficiency. Additionally, the thermal conduction resistance through the spread droplet would be smaller, enhancing evaporative cooling.[65] However, as droplet and characteristic dimensions of the pattern become of the same order, the hydrodynamics are expected to deviate from those observed here on the patterned surfaces. For very small spot sizes and/or small inter-pattern distances, satellite droplets would likely not form, as surface tension forces would prevent droplet break-up.[28] Thus, for microscopic droplets, we would expect to only observe trends in cooling efficiency similar to those of superhydrophobic or sticky hydrophobic surfaces, but not wettability-patterned surfaces, as shown in Fig. 6b for millimetric droplets.

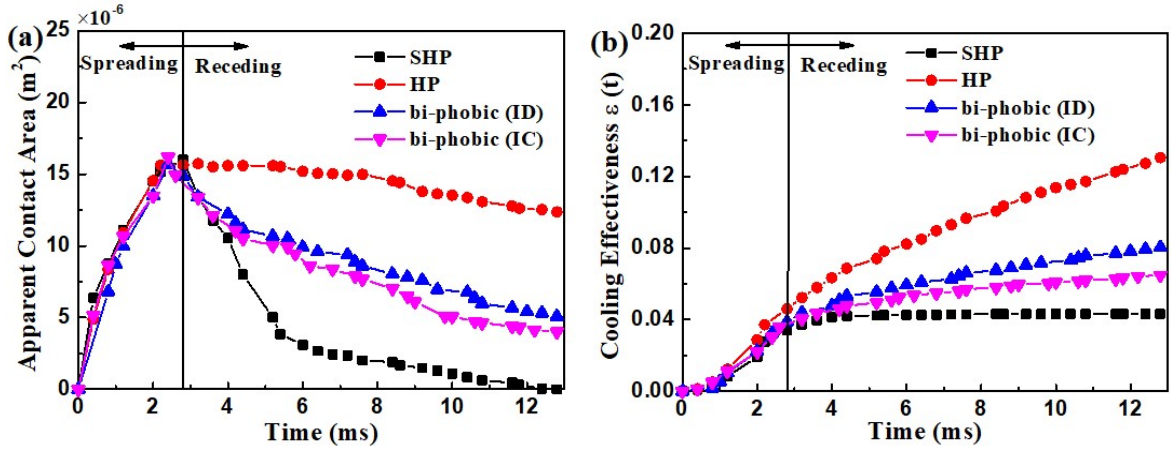

**FIG. 6.** Transient behavior of (a) apparent contact area and (b) cooling effectiveness for different surfaces.

### D. Influence of evaporation

In addition to sensible heat transfer (*i.e.*, the change in droplet or substrate temperature), which correlates to the contact area, we also observe enhanced evaporation, as evidenced by the local maxima in heat fluxes at the TPCL in Figs. 4 and 7.[48,66,67] To analyze the influence of the contact line and its dynamics on heat transfer, we plot the total heat transfer of the contact line region at maximum spreading and maximum recoil (12.8 ms) in Fig. 7a. We also show the temperature (top) and heat flux (bottom) distributions of the different surfaces at maximum spreading (Fig. 7b) and maximum recoil (Fig. 7c), respectively. The TPCL region is defined as ±3 pixel surrounding the white dotted lines in Fig. 7b,c. While the temperature exhibits a minimum in the central region of the droplet, *i.e.* the area of longest liquid-solid contact time, we observe the highest heat fluxes predominantly near the TPCL. This is in good agreement with the results of Karchevsky *et al.*,[67] who concluded that the heat flux in the region of the contact line exceeds the average heat flux of the entire droplet by a factor of 5-7. At maximum spreading, the hydrophobic surface exhibits the highest heat transfer rate near the contact line region, as seen in Fig. 7a. This is attributed to the surface roughness and the droplet's Wenzel state, which increase the effective contact area on the contact line region, leading to faster heat transfer as compared to other surfaces. Due to pinning and hydrodynamic oscillations at later stages of the impact process, which transport colder water from the bulk droplet to the interface, the hydrophobic surface is the only one that has areas of high local heat flux within the center region of the droplet. The supplementary Video S2 visualizes this phenomenon. Droplet oscillation usually occurs when a droplet is pinned and the initial impact kinetic energy is dissipated through viscous dissipation,[68] and typically results in more non-uniform heat transfer, but with lower overall heat flux values.[17] Hence the highest heat flux does not occur around the contact line region at maximum recoil for the hydrophobic surface, as can be seen when comparing Fig. 7b and Fig. 7c. In all other cases, contact line heat transfer becomes increasingly important, and is most obvious at the maximum recoil stage of the two bi-phobic surfaces. As illustrated in Fig. 7a, the TPCL heat transfer on the bi-phobic surfaces is not significantly lower than on the superhydrophobic surface at maximum spreading. However, the bi-phobic surface has a significantly higher heat transfer at maximum recoil; even compared to the sticky hydrophobic surface. Compared to the hydrophobic surface, the TPCL heat transfer increased by 17 to 27% for the bi-phobic surfaces. The increased contact line length from the satellite droplets increases the evaporation rate, which increases the importance of the contact line heat transfer to the overall heat exchange.[69] The higher contact line heat transfer on the ID surface is in line with the higher trapped liquid volume (approximately 9.5% of the initial volume) compared to the IC bi-phobic surface (7.2%), leading to a longer contact line. As noted earlier, the improved wetting on the hydrophobic spots also accounts for enhanced heat transfer, as can be seen for the bi-phobic surfaces at maximum spreading in Fig. 7b.

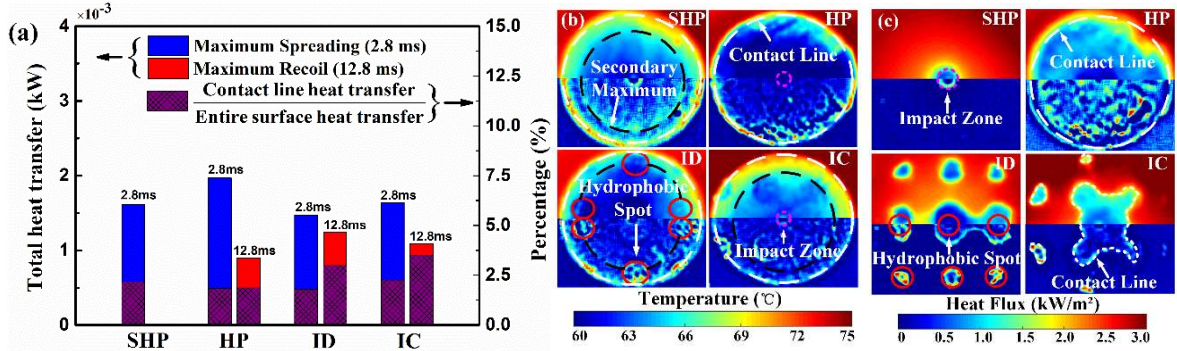

FIG. 7. (a) Total heat transfer of the contact line region (±3 pixel surrounding the dashed white lines in (b) and (c)) at maximum spreading (2.8 ms) and recoil (12.8 ms), and heat flux and temperature distributions at (b) maximum spreading and (c) maximum recoil for the different surfaces, with the surface temperature in the upper panel and the heat flux distribution in the lower panel.

### E. Influence of impact location

In Fig. 7b we can also distinguish an area of different temperature and heat flux in the center of the droplets. As mentioned above, the location of impact influences the dynamics of the impingement process.[44] Figure 8 highlights the optical and thermal signatures of a droplet on an IC bi-phobic surface, *i.e.* where the center of droplet impact is located on the superhydrophobic matrix. It can be seen that the impact zone has a higher temperature and lower heat flux compared to the area surrounding it. Unfortunately, due to the limited spatial resolution of the IR camera, we are unable to resolve the exact reason for this anomaly. Two explanations are possible. In the optical image, the impact zone appears darker, which suggests the imbibition of liquid into the nanostructure due to the water hammer and shock wave pressure during the early stages of impact.[70–72] The improved contact between the liquid and the solid then leads to a better heat transfer during the early stages of spreading (not shown here), and consequently higher droplet temperature at later stages (shown here at maximum spreading). This relatively higher droplet temperature decreases the driving potential for heat transfer, as exemplified by the vanishing heat flux in the bottom right panel in Fig. 8. On the other hand, droplet impact can also lead to the entrapment of a central air bubble.[73–76] The high thermal resistance of the trapped bubble at the impact zone would decrease the conduction rate from the heated substrate to the bulk droplet (hence the reduced heat flux), resulting in an elevated interfacial temperature.[77] Further studies at higher spatial resolution are required to unveil the exact mechanism for this impact-zone anomaly.

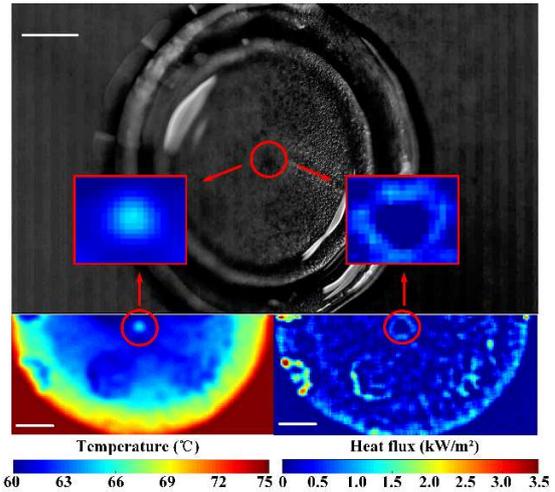

FIG. 8. Images of the impact zone at maximum spreading (2.8 ms) on the IC bi-phobic surface. Scale bars are 0.8 mm.

### IV. CONCLUSIONS AND OUTLOOK

In this study we have compared the dynamic wetting and heat transfer behavior during droplet impingement onto heated superhydrophobic, hydrophobic, and bi-phobic surfaces. We can draw the following conclusions:
1) As expected, all surfaces show the same hydrodynamic and thermal signatures during droplet spreading, while stark differences appear

during receding. Due to pinning on the wettability-patterned surface, small satellite droplets are formed, which effectively increases the contact area and contact line length compared to superhydrophobic surfaces, while decreasing the probability of surface flooding. The existence of satellite droplets increases the contribution of evaporative cooling on the surfaces.
2) The heat transfer performance has been found to be substantially different during the different phases of the droplet impingement process. The maximum heat flux occurs before the droplet reaches its maximum extent, while it is nearly constant during receding. During spreading, the heat flux distribution has two maxima, mapping the hydrodynamic flow inside the droplets. At maximum spreading and during receding, heat transfer at the contact line becomes increasingly important.
3) While on the superhydrophobic surface the majority of heat transfer (≈80%) occurs during spreading, on bi-phobic surfaces spreading and receding contribute equally to the total heat transfer. This can, again, be attributed to the enhanced evaporation at the periphery of the satellite droplets on the hydrophobic spots.

Findings from this work have the potential to improve the efficiency of thermal management applications or of micro-reactors in chemical and biomedical engineering. More work is needed to fully understand the influence of the central impact region (bubble or imbibition), as well as to uncover the spatial and temporal limitations. For example, from a thermal perspective, reducing the size and spacing between hydrophobic spots would be beneficial to enhancing heat transfer even more by increasing the contact line length and decreasing the conduction resistance through the small satellite droplets. If small enough, evaporation times of satellite droplets could be of the same order as the time interval between successive impact events, increasing the total cooling effectiveness while mitigating flooding. Preliminary results on multi-droplet impact are presented in section E of the SI. However, hydrodynamically, an infinitely small spacing or very fine patterns would not lead to the rupture of liquid bridges and the formation of satellite droplets. Similarly, there exists a trade-off between droplet size, maximum spreading diameter, and the characteristic dimensions of the bi-phobic pattern. The findings presented in this work are expected to apply to cases where the maximum spreading diameter is bigger than spot size and spacing. For much smaller droplets, such as during spray cooling, the influence of substrate wettability patterning would need to be re-evaluated.

## SUPPLEMENTARY MATERIAL

Supplementary Information (PDF)
Video S1: SHP - Impact height 80mm - Temperature 60°C. (AVI)
Video S2: HP - Impact height 40mm - Temperature 45°C. (AVI)
Video S3: ID - Impact height 80mm - Temperature 60°C. (AVI)
Video S4: IC - Impact height 120mm - Temperature 75°C. (AVI)

## ACKNOWLEDGEMENTS

The authors would like to thank Jianxing Sun for providing scanning electron microscopy images of the Glaco layer on a glass substrate. The authors acknowledge financial support from Washington University in St. Louis and the Institute of Materials Science and Engineering for the use of instruments and staff assistance. W. L. Qi acknowledges financial support from the China Scholarship Council (CSC).

## AIP PUBLISHING DATA SHARING POLICY

The data that support the findings of this study are available from the corresponding author upon reasonable request.

# Supporting Information to:

# Dynamic wetting and heat transfer during droplet impact on bi-phobic wettability-patterned surfaces

Wenliang Qi(齐文亮),[1] and Patricia B. Weisensee[1,2,a]

[1] Department of Mechanical Engineering & Materials Science, Washington University in St. Louis, Missouri 63130, USA
[2] Institute of Materials Science and Engineering, Washington University in St. Louis, Missouri 63130, USA

[a] Corresponding author E-mail: p.weisensee@wustl.edu

The objective of the supporting information is to provide more analysis on dynamic wetting and heat transfer during droplet impact at different impact conditions, and to demonstrate the universality of the results presented in the main manuscript. Transient behaviors of contact line velocity, average surface temperature, and total heat flux were analyzed in this supporting information. We conducted additional experiments at impact heights = 40±1 mm, 80±1 mm, and 120±1 mm, leading to impact velocities of 0.52 m/s, 0.74 m/s, and 0.96 m/s, respectively, with Weber numbers 5, 10, and 17, respectively. The surfaces were heated to 45±0.5°C, 60±0.5°C, and 75±0.5°C prior to impact.

## A. Uncertainty Analysis

Table SI summarizes the relevant parameter and their associated percentage uncertainty. All listed values are to a 97% confidence level. The heat conduction term (eq. 4 in the main manuscript) is extremely sensitive to the spatial signal noise of the input temperature field. Since only five different pixel elements (the one of interest and 4 neighboring) are used for the calculation of the surface heat flux, it is very prone to noise. We applied a Gaussian filter to smoothen the temperature signal and suppress its sensitivity on the heat flux distribution. The size of the filter was $n = 5$ and the standard deviation was $\sigma_n = 2.5$. The use of the Gaussian filter, the energy balance to determine the electrical resistance, and the experimental uncertainties listed in table S1 result in an uncertainty in the reported heat flux values of 4.4 to 11%.

TABLE S1. Experimental uncertainty

| Parameter | | [%] |
|---|---|---|
| Substrate temperature, $T_s$ | ± | 4.1 |
| Ambient temperature, $T_{air}$ | ± | 0.04 |
| Contact angle, $\theta$ | ± | 1.2 |
| Droplet diameter, $D$ | ± | 2.2 |
| Impact height, $H$ | ± | 2.5 |
| Electrical voltage, $V_{Cr}$ | ± | 2.7 |
| Electrical current, $I_s$ | ± | 3.2 |

To ensure good reproducibility of the results, we repeated all experiments at least three times. Figure S1 exemplarily compares average temperature and total integrated heat flux on IC bi-phobic surfaces for the three impact events. The relative error between them is less than 10%, and hence falls within the general experimental uncertainty.

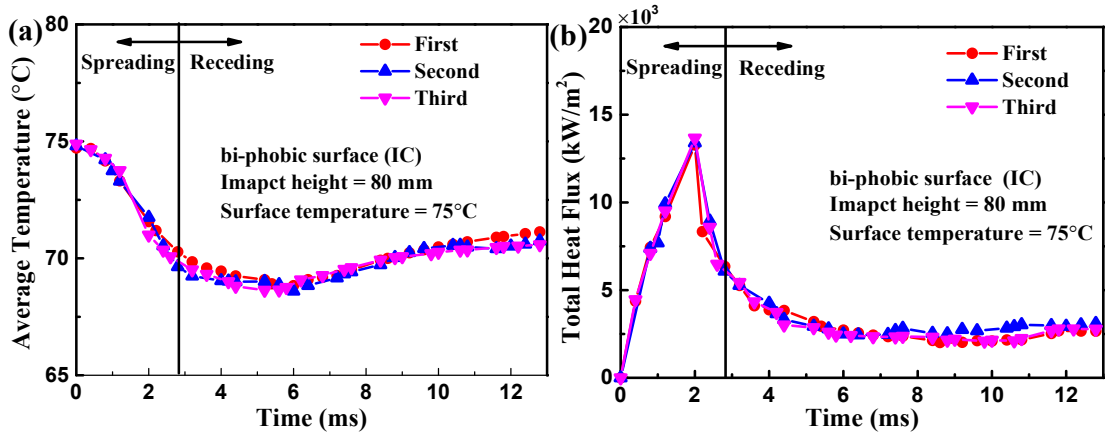

FIG. S1. Comparison of the transient behavior of (a) average temperature and (b) total integrated heat flux for three equivalent impact conditions on IC bi-phobic surfaces.

## B. Influence of surface temperature and impact height

Figure S2 quantifies the influence of surface temperature and impact height on the spreading dynamics on the different surfaces. As seen in Fig. S2a, surface temperature has a minimal effect on the maximum spreading diameter and spreading time, in good agreement with previous studies.[1,2] Since spreading is inertia-dominated, a slight change in viscosity and surface tension with temperature does not influence its dynamics.[3] Similarly, changes in surface wettability play only a minor role, since for all surfaces $\theta_A \geq 90°$. This result is in good agreement with previous experimental and numerical findings for droplet impact and spreading on textured and wettability-patterned surfaces.[4–6] As the droplet spreads over the textured surface, entrapped air inside the porosity cannot escape quickly enough, leading to a cushioning effect. In our experiments, all samples have the same surface texture, with only surface chemistry varying. As such, air drainage from the porosity is expected to be similar on all surfaces, hence leading to the same amount of viscous dissipation, irrespective of the static wettabilities of the different surfaces. As expected,[7] maximum spreading diameters increase with increasing impingement height, while spreading times remain unchanged, as shown in Fig. S2b. Spreading times scale with the capillary-inertial time scale, which is independent of impact velocity, whereas maximum spreading is dictated by an energy or momentum balance before impact and at maximum spreading (there is an ongoing debate on the physics determining the maximum spreading, however, it is beyond the scope of this work to elucidate the exact mechanism dictating spreading dynamics). After reaching their maximum spreading diameter, droplets either get pinned (HP), or recede (SHP, ID, IC). Receding dynamics are discussed in detail in the main manuscript. Figure S3 shows the detailed optical and thermal information for impact from 120 mm to supplement Fig. 4 from the main manuscript (80 mm).

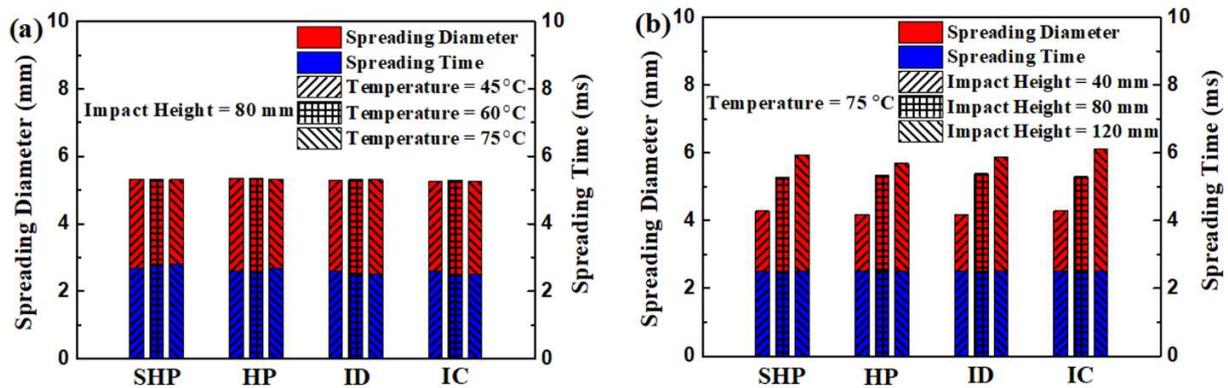

FIG. S2. Spreading diameter (red) and corresponding spreading time (blue) for different surfaces. (a) Influence of surface temperature. (b) Influence of impact height.

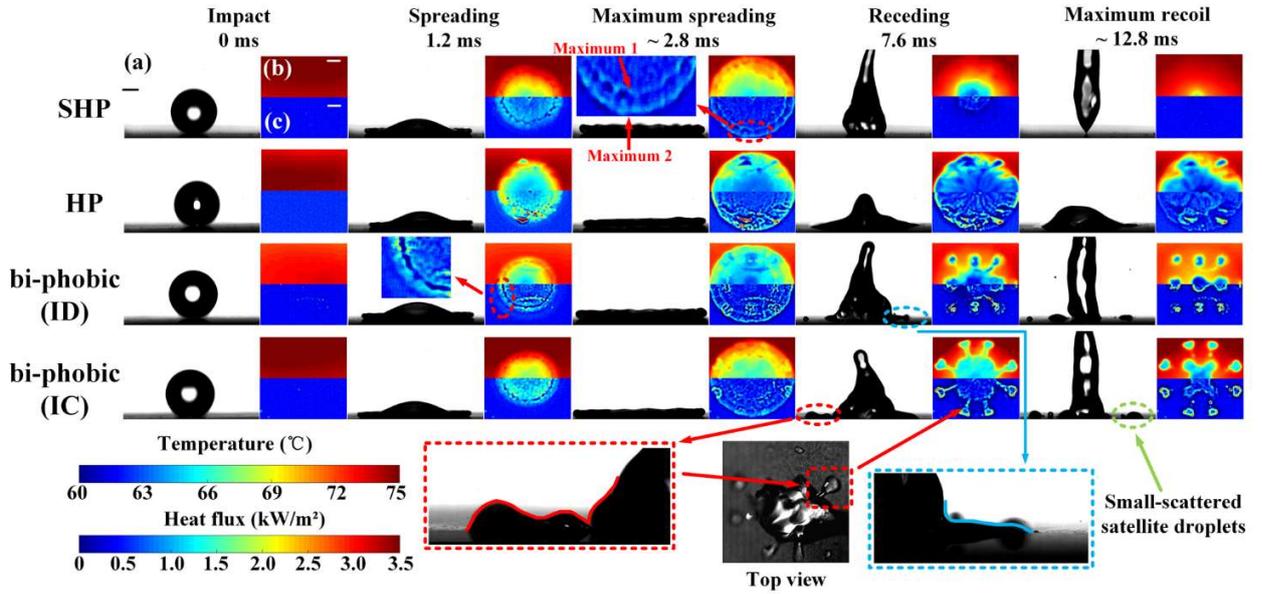

**FIG. S3.** Optical and thermal images for impact from 120 mm (v = 0.74 m/s), accompanying Fig. 4 from the main manuscript. Some of the features as more pronounced at this impact velocity. (a) High-speed camera images of droplets and (b) corresponding temperature and (c) heat flux distributions for the different surfaces. Scale bars are 0.8 mm.

## C. Evaluation of the contact line velocity

The temporal evolution of the contact line velocity on different surfaces is shown in Fig. S4. It can be seen that all surfaces have the same contact line velocity in the inertia-dominated spreading phase, consistent with the same spreading times and spreading diameters shown in Fig. S2.[2] However, the different dynamics during receding become obvious in this figure. Due to viscous dissipation and a high wetting force in the Wenzel state, droplets remained pinned on the hydrophobic surface after reaching maximum spreading, leading to a receding contact line velocity of 0. On the other surfaces, the stored surface energy gets converted back into kinetic energy,[2] leading to a reversal in the spreading direction and an increase in contact line velocity (absolute value). On the superhydrophobic surface, the velocity reaches its maximum just before the droplet elongates upwards. The initial temporal evolution of the contact line velocity on the bi-phobic surfaces is similar. However, when the contact line recedes over the hydrophobic spots, some kinetic energy is dissipated by the non-uniform frictional resistance, while neighboring liquid parcels continue to recede on the superhydrophobic matrix.[8] Eventually, small satellite droplets break off, leading to a spike in contact line velocity. This local increase in contact line velocity occurs first on the IC, then the ID surface, due to the geometry of the patterning.

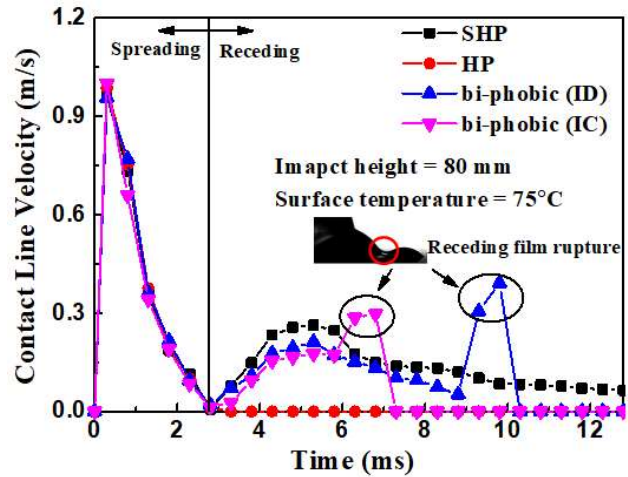

FIG. S4. Transient behavior of the absolute value of the contact line velocity for different surfaces. Note that the contact line velocity on bi-phobic surface is not symmetric due to the different surface wettability. This figure shows the contact line velocity of the droplet in a side view parallel to the hydrophobic dots.

**D. Influence of surface temperature and impact height**

Figure S5 shows the temporal evolution of the average temperature and total heat flux under different impact height and surface temperature conditions. Several general trends can be observed in this figure. First, the heat transfer performance is substantially different during the different phases of the drop impingement process. Second, and as already shown in the main manuscript, the minimum average temperature occurs during the receding phase (after ≈ 3-4 ms), followed by a relatively long period of modest increase, irrespective of surface wettability, impact height, or surface temperature. The average temperature increases with time due to the decreasing contact area between droplet and surface, which restricts the heat transfer between surface and droplet. Third, the total heat transfer peaks before maximum spreading, and becomes approximately constant during the receding phase, again irrespective of surface and impact conditions. As the impact height increases (Fig. S5a,b), the spreading diameter and the available area for heat transfer also increase, leading to enhanced conduction heat transfer between droplet and susbtrate.[9] Moreover, a higher impact height results in a thinner film, which equally affects the cooling of the surface. As expected, the average temperature and total heat flux increase with increasing initial surface temperatures, as shown in Fig. S5c,d. The higher temperature difference between droplet and substrate leads to a) a higher driving potential for heat transfer, an b) an increased total available amount of thermal energy (~ $c_p \Delta T$). By comparing the results of impact height and surface temperature, we can conclude that the impact height plays a minor role in influencing heat transfer during the impact process.

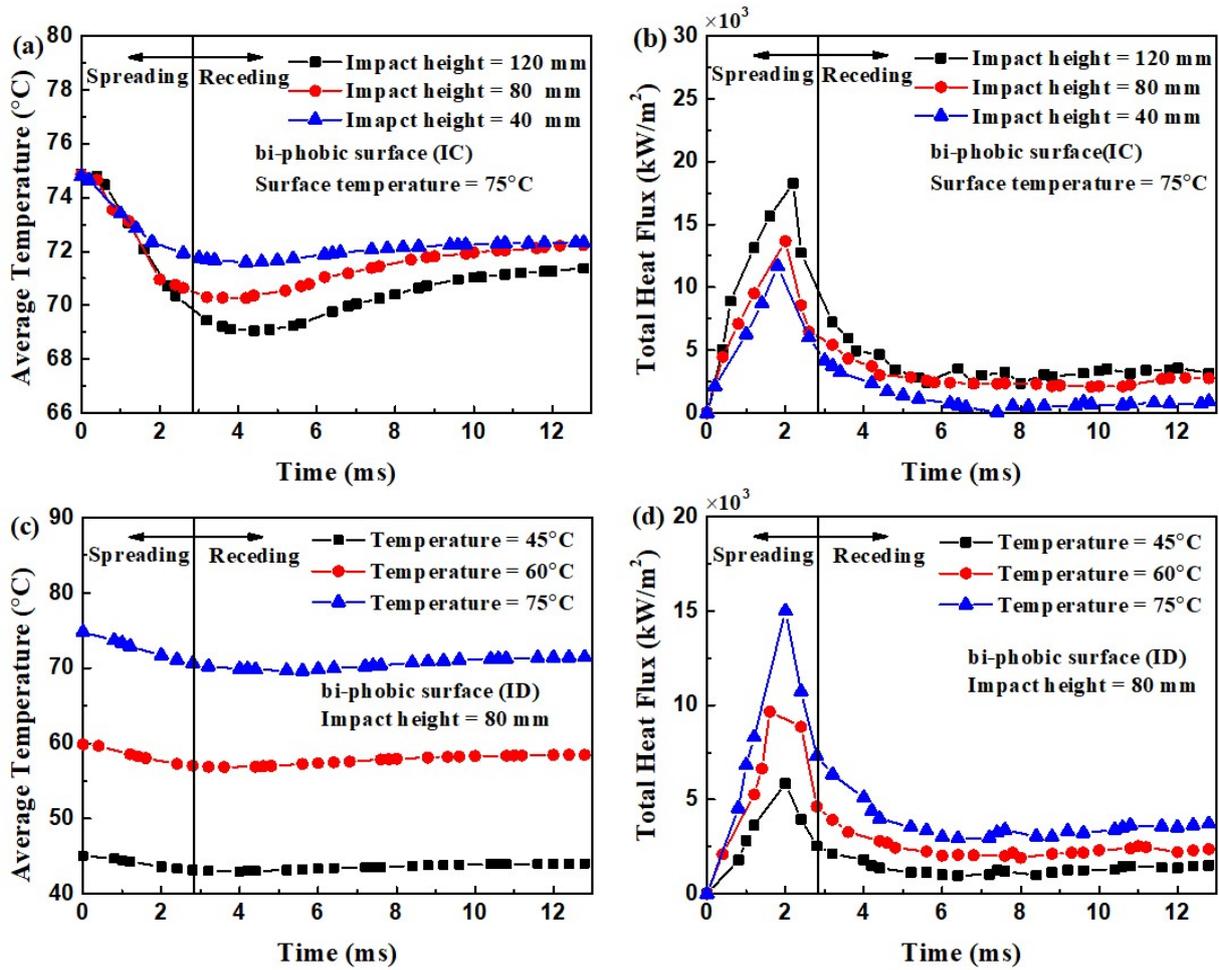

FIG. S5. Transient behavior of average temperature (a,c) and total heat flux (b,d) for the two bi-phobic surfaces (IC: top, ID: bottom) for different experiment conditions (varying impact height: top, varying surface temperature: bottom).

### E. Influence of multi-droplet impact

In an effort to evaluate the benefit of using wettability-patterned surfaces for heat transfer applications with successive droplet impact events, such as during spray cooling, we studied the average substrate temperature evolution for five droplets with a time delay of approximately 250 ms between impact events. The number of impact events was limited due to experimental constraints, as measurements were only possible for a horizontally mounted substrate. Over time, even on the superhydrophobic surface, bouncing droplets would re-deposit on the surface and eventually lead to flooding. We hypothesized that - while the hydrophobic sample has the highest cooling capacity for a single droplet - the bi-phobic surfaces would have a superior performance for a long-term operation on a tilted surface, as they can mitigate flooding. Unfortunately, five impact events on a horizontal sample were not sufficient to fully confirm this hypothesis, as seen in Fig. S6. We can nonetheless draw some conclusions. Obviously, the superhydrophobic surface has close to no cooling effect beyond the actual duration of contact (~ 10 ms), whereas the temperature on the other surfaces decreases continuously, with the hydrophobic surface having the strongest cooling effect. The two bi-phobic surfaces perform equally well (or bad). It is interesting to note, though, that the cooling effectiveness, *i.e.* temperature drop, for each new impact event decreases drastically on the hydrophobic surface as a liquid puddle builds up, as seen in the inset in Fig. S6. The successive cooling effectiveness on the bi-phobic surfaces deceases much slower ($\Delta T/\Delta n \approx$ -0.2 K/impact for the bi-phobic vs. $\Delta T/\Delta n \approx$ -0.3 K/impact for the hydrophobic surface). This suggests that for a higher number of impacting droplets, the wettability-patterned surface can potentially outperform the single-wettability surface due to the prevention of flooding (*vs.* hydrophobic or hydrophilic) and/or the addition of pinned satellite droplets which contribute to enhanced evaporative cooling (*vs.* superhydrophobic). The second interesting observation concerns the temperature evolution in-between impact events. While on the hydrophobic surface the temperature monotonically increases, the temperature on the bi-phobic surfaces first sharply increases when the bulk droplet lifts off, but then continuously decreases until the next impact event. This observations again substantiates the importance of the evaporation of the satellite droplets on the overall cooling efficiency. By decreasing the size of the hydrophobic spots, decreasing the distance between them, increasing the time period between successive droplet impact events, and tilting the surface, we would expect the bi-phobic surfaces to out-perform the hydrophobic surface. Furthermore, as we have shown recently, evaporation on, for example, a vertical substrate could additionally enhance the evaporation heat transfer by influencing convective currents inside and outside of the droplet.[10]

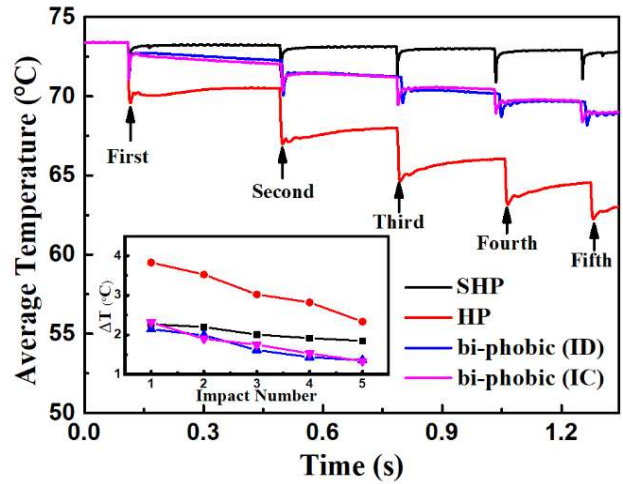

**FIG. S6.** Average temperature of the heater surface for successive droplet impingements on the different surfaces for an impact height of 210mm and an initial surface temperature of 73.5° C. Inset: Variation of surface temperature at the moment of droplet impact with the droplet impact number.